\documentclass{article}%
\usepackage{amsmath}
\usepackage{amsfonts}
\usepackage{amssymb}
\usepackage{graphicx}%
\newcommand{\bpartial}{\mathop{\partial\kern -4pt\raisebox{.8pt}{$|$}}}
\begin{document}

\title{ Superposition Principle and the Problem of the Additivity of the Energies and
Momenta of Distinct Electromagnetic Fields}
\author{Eduardo Notte-Cuello$^{(1)}$ and Waldyr A. Rodrigues Jr.$^{(2)}$\\$^{(1)}${\small Departamento de} {\small Matem\'{a}ticas,}\\{\small Universidad de La Serena}\\{\small Av. Cisternas 1200, La Serena-Chile}\\$^{(2)}\hspace{-0.1cm}${\footnotesize Institute of Mathematics, Statistics and
Scientific Computation}\\{\footnotesize IMECC-UNICAMP CP 6065}\\{\footnotesize 13083-859 Campinas, SP, Brazil}\\{\small e-mail:} {\small enotte@userena.cl and walrod@ime.unicamp.br}}
\maketitle

\begin{abstract}
In this paper we prove in a rigorous mathematical way (using the Clifford
bundle formalism) that the energies and momenta of two \textit{distinct} and
\textit{arbitrary} free Maxwell fields (of finite energies and momenta) that
are superposed are additive and thus that there is \textit{no} incompatibility
between the principle of superposition of fields and the principle of
energy-momentum conservation, contrary to some recent claims. Our proof
depends on a noticeable formula for the energy-momentum densities, namely,
Riesz formula $\star\mathcal{T}^{\mathbf{a}}=\frac{1}{2}\star(F\theta
^{\mathbf{a}}\tilde{F})$, which is valid for any electromagnetic field
configuration $F$ satisfying Maxwell equation ${\mbox{\boldmath$\partial$}}%
F{=0}$.

\end{abstract}

\section{Introduction}

In this paper we analyze the compatibility of the principle of superposition
for Maxwell fields and the principle of energy-momentum conservation. More
exactly, we prove in a rigorous mathematical way, using the Clifford bundle
formalism, that the energy and momentum of two distinct and otherwise
arbitrary Maxwell fields (of finite energy) are additive. Let us now describe
precisely our problem.

In a given inertial frame $\mathbf{I}=\partial/\partial x^{0}\in\sec TM$ in
Minkowski spacetime (see Appendix for some details) with a natural adapted
coordinates $\{x^{\mu}\}$ in Einstein-Lorentz-Poincar\'{e} gauge\footnote{We
use units where the velocity of light $c$ has the numerical value $1$ and so
the timelike coordinate $x^{0}=t$.} we have two identical antennas that are
put on at time $t=-\tau$ and which are able to produce two distinct
electromagnetic fields, which we take for simplicity, as being of the same
time duration $\tau$. So, at time $t=0$ we have two electromagnetic field
configurations denoted $F_{1}(0,\mathbf{x})$ and $F_{2}(0,\mathbf{x})$, which,
of course, have \textit{compact support} in a region $\mathfrak{R\subset}$
$\mathbb{R}^{3}$ (the rest space of the inertial frame) and which we suppose
are moving in opposite directions (to fix the ideas the $z$-direction). We
suppose moreover that the two antennas are separated by a distance $D>\tau$,
which means that one does not affect the other during the time they are
generating the \ arbitrary electromagnetic field configurations $F_{1}%
(0,\mathbf{x})$ and $F_{2}(0,\mathbf{x})$. These fields can then be taken as
Cauchy data for Maxwell equations and at any time $t$ satisfy Maxwell
equation\footnote{No misprint here. Maxwell \textit{equation} in free space
${\mbox{\boldmath$\partial$}}F=0$ is the equation of motion for an
electromagnetic field configuration $F\in\sec\bigwedge\nolimits^{2}T^{\ast
}M\hookrightarrow\sec\mathcal{C\ell(}M,\mathtt{\eta})$, where $\mathcal{C\ell
(}M,\mathtt{\eta})$ is the Clifford bundle of differential forms. See the
Appendices for explanation of the symbols and the main definitions and
\cite{rodcap} for a detailed exposition.} in free space%
\begin{equation}
{\mbox{\boldmath$\partial$}}F_{1}=0,\text{ }{\mbox{\boldmath$\partial$}}%
F_{2}=0. \label{em6}%
\end{equation}

$\ $Let $d$ be the distance between the wave fronts of the two pulses at $t=0$
measured along the $z$ axis. Now, at time $\mathfrak{t}=(d+\tau)/2$ the pulses
$F_{1}(\mathfrak{t},\mathbf{x})$ and $F_{2}(\mathfrak{t},\mathbf{x})$ (which
will be always diffracted\footnote{This is a consequence of the non focusing
theorem. See \cite{wu} and also \cite{farlow}.} in relation to the initial
configurations $F_{1}(0,\mathbf{x})$ and $F_{2}(0,\mathbf{x})$) which move
with the velocity of light $c=1$ fill the same region of space and generate a
total electromagnetic field
\begin{equation}
F(\mathfrak{t},\mathbf{x})=F_{1}(\mathfrak{t},\mathbf{x})+F_{2}(\mathfrak{t}%
,\mathbf{x}), \label{sup1}%
\end{equation}
which satisfy also the free Maxwell equation (with appropriate initial
conditions) due to the principle of superposition valid for linear partial
differential equations,%
\begin{equation}
{\mbox{\boldmath$\partial$}}F=0. \label{maxwell1}%
\end{equation}

In the Clifford bundle formalism \ the energy-momentum densities
$\star\mathcal{T}^{\mathbf{a}}\in\sec\bigwedge\nolimits^{3}T^{\ast
}M\hookrightarrow\sec\mathcal{C\ell(}M,\mathtt{\eta})$ $(\mathbf{a}=0,1,2,3)$
of an electromagnetic field configuration $F$ is given by Riesz
formula\footnote{Notice that $\mathcal{T}_{\mathbf{a}}=\mathcal{T}%
_{\mathbf{ab}}\theta^{\mathbf{b}}$, where $\mathcal{T}_{\mathbf{ab}}$ are the
components of the \textit{usual} \cite{parrot} energy-momentum tensor of
Maxwell theory. Discussions about the appropriateness for the use of the
\textit{usual }energy-momentum tensor for the description of energy-momentum
propagation are given in \cite{parrot,ribaric}.} (see Appendix)
\begin{equation}
\star\mathcal{T}^{\mathbf{a}}=\frac{1}{2}\star(F\theta^{\mathbf{a}}\tilde{F}),
\label{ta}%
\end{equation}
and the energy-momentum $P_{F}^{\mathbf{a}}$ of the field configuration at
time $t=\mathfrak{t}$ is given by
\begin{equation}
P_{F}^{\mathbf{a}}=\int\nolimits_{B_{2}}\star\mathcal{T}^{\mathbf{a}},
\label{pa}%
\end{equation}
where\footnote{See Figure 1 for the defintion of regions $B_{1}$, $B_{2}$,
$B_{3}$, $B_{3}^{\prime}$, $C_{1}$,$C_{2},C_{1}^{\prime}$ and $C_{2}^{\prime}%
$.} $B_{2}$ \ is contained in the constant time hypersurface $t=$
$\mathfrak{t}$ in Minkowski spacetime.

Now, due to Eq.(\ref{sup1}) we have%

\begin{align}
\star\mathcal{T}^{\mathbf{a}}  &  =\frac{1}{2}\star(F\theta_{\mathbf{a}}%
\tilde{F})=\frac{1}{2}\star(\left(  F_{1}+F_{2}\right)  \theta^{\mathbf{a}%
}\widetilde{\left(  F_{1}+F_{2}\right)  })\nonumber\\
&  =\frac{1}{2}\star(F_{1}\theta^{\mathbf{a}}\tilde{F}_{1}+F_{2}%
\theta^{\mathbf{a}}\tilde{F}_{2}+F_{1}\theta^{\mathbf{a}}\tilde{F}_{2}%
+F_{2}\theta^{\mathbf{a}}\tilde{F}_{1})\label{em10}\\
&  =\star\mathcal{T}_{1\mathbf{\ \ }}^{\mathbf{a}}+\star\mathcal{T}%
_{2\mathbf{\ \ }}^{\mathbf{a}}+\star\mathcal{K}^{\mathbf{a}}\nonumber
\end{align}
where
\begin{align}
\star\mathcal{T}_{1\mathbf{\ \ }}^{\mathbf{a}}  &  =\frac{1}{2}\star
(F_{1}\theta^{\mathbf{a}}\tilde{F}_{1}),\text{ }\star\mathcal{T}%
_{2\mathbf{\ \ }}^{\mathbf{a}}=\frac{1}{2}\star(F_{2}\theta^{\mathbf{a}}%
\tilde{F}_{2}),\nonumber\\
\star\mathcal{K}^{\mathbf{a}}  &  =\frac{1}{2}\star(F_{1}\theta^{\mathbf{a}%
}\tilde{F}_{2}+F_{2}\theta^{\mathbf{a}}\tilde{F}_{1}). \label{em11}%
\end{align}
Then we have that
\begin{equation}
P_{F}^{\mathbf{a}}=\int\nolimits_{B_{2}}\star\mathcal{T}^{\mathbf{a}}%
=\int\nolimits_{B_{2}}\star\mathcal{T}_{1}^{\mathbf{a}}+\int\nolimits_{B_{2}%
}\star\mathcal{T}_{2}^{\mathbf{a}}+\int\nolimits_{B_{2}}\star\mathcal{K}%
^{\mathbf{a}}. \label{em12}%
\end{equation}

We want to prove that the energy and momentum of the field configuration
$F_{1}$ and $F_{2}$ \ at time $t=\mathfrak{t}$ (and indeed at any time) is
additive, i.e., is given by
\begin{equation}
P_{F}^{\mathbf{a}}=P_{F_{1}}^{\mathbf{a}}+P_{F_{2}}^{\mathbf{a}},\text{
}\mathbf{a}=0,1,2,3, \label{thesis}%
\end{equation}
with%
\begin{equation}
P_{F_{1}}^{\mathbf{a}}=\int\nolimits_{B_{2}}\star\mathcal{T}_{1}^{\mathbf{a}%
},\text{ }P_{F_{2}}^{\mathbf{a}}=\int\nolimits_{B_{2}}\star\mathcal{T}%
_{2}^{\mathbf{a}}. \label{em13}%
\end{equation}

This problem is a nontrivial one, and has been not discussed in the literature
in an appropriate and satisfactory way according to our view. For example, in
\cite{levine} (written in 1980) the author said that he found the problem
discussed in only two (\cite{jenkins,strong}) out of 50 textbooks he has
examined. Moreover, from a few papers published in the literature, we found
some good ideas, but none offers a rigorous solution for the problem. Worse,
some papers and books \cite{cornille,ku,ku1} have very odd \ and/or dubious
statements. Indeed, in \cite{cornille} it is said that Eq.(\ref{em12}) implies
in \textit{non} conservation of energy-momentum. The statement about non
conservation of energy-momentum is also done by the author of \cite{ku,ku1}
who says that results of recent experiments \cite{kuetal} endorse his
statement\footnote{A more intelligible and realistic analysis of light
transmission through two slits is given in \cite{welti}.}. \ In
\cite{klein-furtak} the double slit interference with monochromatic
waves\footnote{Which, of course, do not have compact support in $\mathbb{R}%
^{3}.$} is analyzed and it is said that energy-momentum is conserved only
\textit{after }spatial average. On the other hand, e.g., \cite{gauthier}
\ shows that Eq.(\ref{thesis}) is the correct one for the case of two
\textit{plane} waves moving in opposite directions, but since waves of this
kind (which do not have compact support) have infinite energy when the
integration in Eq.(\ref{em12}) is done in all space, his approach cannot in
any way be considered satisfactory. In \cite{chen} it is proposed that
energy-momentum tensors that differ from an exact differential must be
considered equivalent. This is a good idea, if it could be proved (something
which has not been done in \cite{chen}) that
\[
\star\mathcal{K}^{\mathbf{a}}=-d\star\mathcal{E}^{\mathbf{a}}%
\]
for some $\star\mathcal{E}^{\mathbf{a}}\in\sec\bigwedge\nolimits^{2}T^{\ast
}M\hookrightarrow\sec\mathcal{C\ell(}M,\mathtt{\eta})$ which goes to zero at
spatial infinity at time $t=\mathfrak{t}$, since in this case we can write
using Stokes theorem that
\begin{equation}
\int\nolimits_{B_{2}}\star\mathcal{K}^{\mathbf{a}}=-\int\nolimits_{B_{2}%
}d\star\mathcal{E}^{\mathbf{a}}=-\int\nolimits_{\partial B_{2}}\star
\mathcal{E}^{\mathbf{a}}=0 \label{emcrucial}%
\end{equation}
In the Section 2 we show that this is indeed the case for our problem. In
Section 3 we prove that the energies and momenta of two different superposed
electromagnetic field configurations are indeed additive. In Section 4 we
present our conclusions. The paper have 4 Appendices. Appendix A introduces
the concept of Clifford bundles, some important Clifford algebra identities,
the Hodge star operator as an algebraic operation, and the Dirac operator
acting on sections of the Clifford bundle. Appendix B presents Maxwell
equation ${\mbox{\boldmath$\partial$}}F=J$ and the noticeable formula for the
energy-momentum densities $\star\mathcal{T}^{\mathbf{a}}=\frac{1}{2}%
\star(F\theta^{\mathbf{a}}\widetilde{F})$. In Appendix C we describe the main
features of the standard cylinder in Minkowski spacetime need in the
applications of the Stokes theorem in the main text and in Appendix D we
recall for completeness the generalized Green's formula for differential forms.

\section{Proof that $\star\mathcal{K}^{\mathbf{a}}=-d\star\mathcal{E}%
^{\mathbf{a}}$}

From Maxwell theory it follows (see Appendix) that for any free
electromagnetic field configuration that%
\[
\delta\mathcal{T}^{\mathbf{a}}=-{\mbox{\boldmath$\partial$}\lrcorner
}\mathcal{T}^{\mathbf{a}}=0.
\]
\ Since we obviously have ${\mbox{\boldmath$\partial$}\lrcorner}%
\mathcal{T}_{1}^{\mathbf{a}}={\mbox{\boldmath$\partial$}\lrcorner}%
\mathcal{T}_{2}^{\mathbf{a}}=0$, we necessarily must have that
${\mbox{\boldmath$\partial$}\lrcorner}\mathcal{K}^{\mathbf{a}}=0$, i.e.,
\begin{equation}
{\mbox{\boldmath$\partial$}\lrcorner}\frac{1}{2}{(}F_{1}\theta^{\mathbf{a}%
}\tilde{F}_{2}+F_{2}\theta^{\mathbf{a}}\tilde{F}_{1})=0. \label{em11bis}%
\end{equation}
To show that this is indeed the case, first, observe that
\begin{align}
F_{1}\theta^{\mathbf{a}}\tilde{F}_{2}  &  =\left\langle F_{1}\theta
^{\mathbf{a}}\tilde{F}_{2}\right\rangle _{1}+\left\langle F_{1}\theta
^{\mathbf{a}}\tilde{F}_{2}\right\rangle _{3},\\
F_{2}\theta^{\mathbf{a}}\tilde{F}_{1}  &  =\left\langle F_{2}\theta
^{\mathbf{a}}\tilde{F}_{1}\right\rangle _{1}+\left\langle F_{2}\theta
^{\mathbf{a}}\tilde{F}_{1}\right\rangle _{3}.\nonumber
\end{align}
When the previous equations are added, the terms $\left\langle F_{1}%
\theta^{\mathbf{a}}\tilde{F}_{2}\right\rangle _{3}$ and $\left\langle
F_{2}\theta^{\mathbf{a}}\tilde{F}_{1}\right\rangle _{3}$ cancel and we have
\begin{equation}
F_{1}\theta^{\mathbf{a}}\tilde{F}_{2}+F_{2}\theta^{\mathbf{a}}\tilde{F}%
_{1}=\left\langle F_{1}\theta^{\mathbf{a}}\tilde{F}_{2}+F_{2}\theta
^{\mathbf{a}}\tilde{F}_{1}\right\rangle _{1}.
\end{equation}
Returning to Eq.(\ref{em11bis}) we see that we need only to calculate
${\mbox{\boldmath$\partial$}\lrcorner}\left\langle F_{1}\theta^{\mathbf{a}%
}\tilde{F}_{2}+F_{2}\theta^{\mathbf{a}}\tilde{F}_{1}\right\rangle _{1}$. We
have,
\begin{equation}%
\begin{array}
[c]{l}%
{\mbox{\boldmath$\partial$}\lrcorner}\left\langle F_{1}\theta^{\mathbf{a}%
}\tilde{F}_{2}+F_{2}\theta^{\mathbf{a}}\tilde{F}_{1}\right\rangle _{1}\\
=\left\langle {\mbox{\boldmath$\partial$}}\left(  F_{1}\theta^{\mathbf{a}%
}\tilde{F}_{2}\right)  +{\mbox{\boldmath$\partial$}}\left(  F_{2}%
\theta^{\mathbf{a}}\tilde{F}_{1}\right)  \right\rangle _{0}\\
=\left\langle {\theta}^{\mathbf{b}}D_{e_{\mathbf{b}}}\left(  F_{1}%
\theta^{\mathbf{a}}\tilde{F}_{2}\right)  +{\theta}^{\mathbf{b}}%
D_{e_{\mathbf{b}}}\left(  F_{2}\theta^{\mathbf{a}}\tilde{F}_{1}\right)
\right\rangle _{0}%
\end{array}
\end{equation}
or
\begin{equation}%
\begin{array}
[c]{l}%
{\mbox{\boldmath$\partial$}\lrcorner}\left\langle F_{1}\theta^{\mathbf{a}%
}\tilde{F}_{2}+F_{2}\theta^{\mathbf{a}}\tilde{F}_{1}\right\rangle _{1}\\
=\left\langle \left(  {\theta}^{\mathbf{b}}D_{e_{\mathbf{b}}}F_{1}\right)
\theta^{\mathbf{a}}\tilde{F}_{2}+{\theta}^{\mathbf{b}}F_{1}D_{e_{\mathbf{b}}%
}\left(  \theta^{\mathbf{a}}\right)  \tilde{F}_{2}+{\theta}^{\mathbf{b}}%
F_{1}\theta^{\mathbf{a}}D_{e_{\mathbf{b}}}\tilde{F}_{2}\right. \\
+\left.  \left(  {\theta}^{\mathbf{b}}D_{e_{\mathbf{b}}}F_{2}\right)
\theta^{\mathbf{a}}\tilde{F}_{1}+{\theta}^{\mathbf{b}}F_{2}D_{e_{\mathbf{b}}%
}\left(  \theta^{\mathbf{a}}\right)  \tilde{F}_{1}+{\theta}^{\mathbf{b}}%
F_{2}\theta^{\mathbf{a}}D_{e_{\mathbf{b}}}\tilde{F}_{1}\right\rangle _{0}.
\end{array}
\end{equation}
On the other hand, from the Eq. (\ref{em6}) we have
\begin{equation}
{\theta}^{\mathbf{b}}D_{e_{\mathbf{b}}}F_{1}={\mbox{\boldmath$\partial$}}%
F_{1}=0\qquad\text{and\qquad}{\theta}^{\mathbf{b}}D_{e_{\mathbf{b}}}%
F_{2}={\mbox{\boldmath$\partial$}}F_{2}=0 \label{em14}%
\end{equation}
and recalling that $\theta^{\mathbf{a}}=\delta_{\mu}^{\mathbf{a}}dx^{\mu}$ and
$e_{\mathbf{b}}=\delta_{\mathbf{b}}^{\mu}\partial/\partial x^{\mu}$, we have
that $D_{e_{\mathbf{b}}}\theta^{\mathbf{a}}=0$. Then, Eq. (\ref{em11bis}) can
be written as
\begin{equation}
{\mbox{\boldmath$\partial$}\lrcorner}\frac{1}{2}\left\langle F_{1}%
\theta^{\mathbf{a}}\tilde{F}_{2}+F_{2}\theta^{\mathbf{a}}\tilde{F}%
_{1}\right\rangle _{1}=\frac{1}{2}\left\langle {\theta}^{\mathbf{b}}%
F_{1}\theta^{\mathbf{a}}D_{e_{\mathbf{b}}}\tilde{F}_{2}+{\theta}^{\mathbf{b}%
}F_{2}\theta^{\mathbf{a}}D_{e_{\mathbf{b}}}\tilde{F}_{1}\right\rangle _{0}
\label{em15}%
\end{equation}

Now we examine the term $\left\langle {\theta}^{\mathbf{b}}F_{1}%
\theta^{\mathbf{a}}D_{e_{\mathbf{b}}}\tilde{F}_{2}+{\theta}^{\mathbf{b}}%
F_{2}\theta^{\mathbf{a}}D_{e_{\mathbf{b}}}\tilde{F}_{1}\right\rangle _{0}.$
First observe that
\begin{equation}%
\begin{array}
[c]{ll}%
{\theta}^{\mathbf{b}}\left(  F_{1}\theta^{\mathbf{a}}D_{e_{\mathbf{b}}}%
\tilde{F}_{2}\right)  & ={\theta}^{\mathbf{b}}\left\langle F_{1}%
\theta^{\mathbf{a}}D_{e_{\mathbf{b}}}\tilde{F}_{2}\right\rangle _{1}+{\theta
}^{\mathbf{b}}\left\langle F_{1}\theta^{\mathbf{a}}D_{e_{\mathbf{b}}}\tilde
{F}_{2}\right\rangle _{3}\\
& ={\theta}^{\mathbf{b}}\lrcorner\left\langle F_{1}\theta^{\mathbf{a}%
}D_{e_{\mathbf{b}}}\tilde{F}_{2}\right\rangle _{1}+{\theta}^{\mathbf{b}}%
\wedge\left\langle F_{1}\theta^{\mathbf{a}}D_{e_{\mathbf{b}}}\tilde{F}%
_{2}\right\rangle _{1}\\
& +{\theta}^{\mathbf{b}}\lrcorner\left\langle F_{1}\theta^{\mathbf{a}%
}D_{e_{\mathbf{b}}}\tilde{F}_{2}\right\rangle _{3}+{\theta}^{\mathbf{b}}%
\wedge\left\langle F_{1}\theta^{\mathbf{a}}D_{e_{\mathbf{b}}}\tilde{F}%
_{2}\right\rangle _{3}.
\end{array}
\end{equation}
Then%

\begin{align}
\left\langle {\theta}^{\mathbf{b}}F_{1}\theta^{\mathbf{a}}D_{e_{\mathbf{b}}%
}\tilde{F}_{2}\right\rangle _{0}  &  ={\theta}^{\mathbf{b}}\lrcorner
\left\langle F_{1}\theta^{\mathbf{a}}D_{e_{\mathbf{b}}}\tilde{F}%
_{2}\right\rangle _{1}\nonumber\\
&  =\left\langle F_{1}\theta^{\mathbf{a}}D_{e_{\mathbf{b}}}\tilde{F}%
_{2}\right\rangle _{1}\llcorner{\theta}^{\mathbf{b}}\nonumber\\
&  =\left\langle F_{1}\theta^{\mathbf{a}}\left(  D_{e_{\mathbf{b}}}\tilde
{F}_{2}\right)  {\theta}^{\mathbf{b}}\right\rangle _{0}\nonumber\\
&  =\left\langle F_{1}\theta^{\mathbf{a}}\left(  \tilde{F}_{2}\overleftarrow
{{\mbox{\boldmath$\partial$}}}\right)  \right\rangle _{0}%
\end{align}
where we use the symbol $\left(  D_{e_{\mathbf{b}}}\tilde{F}_{2}\right)
{\theta}^{\mathbf{b}}:=\tilde{F}_{2}\overleftarrow
{{\mbox{\boldmath$\partial$}}}.$\ Since $\tilde{F}_{2}\overleftarrow
{{\mbox{\boldmath$\partial$}}}=\widetilde{({\mbox{\boldmath$\partial$}\ }%
F_{2})}={0}$, we have $\left\langle {\theta}^{\mathbf{b}}F_{1}\theta
^{\mathbf{a}}D_{e_{\mathbf{b}}}\tilde{F}_{2}\right\rangle _{0}=0$. Analogously
we get that $\left\langle {\theta}^{\mathbf{b}}F_{2}\theta^{\mathbf{a}%
}D_{e_{\mathbf{b}}}\tilde{F}_{1}\right\rangle _{0}=0$, and thus
\begin{equation}
\left\langle {\theta}^{\mathbf{b}}F_{1}\theta^{\mathbf{a}}D_{e_{\mathbf{b}}%
}\tilde{F}_{2}+{\theta}^{\mathbf{b}}F_{2}\theta^{\mathbf{a}}D_{e_{\mathbf{b}}%
}\tilde{F}_{1}\right\rangle _{0}=0. \label{em15,5}%
\end{equation}
Now, using Eq.(\ref{em15,5}) in Eq. (\ref{em15}) we have
\begin{equation}
{\mbox{\boldmath$\partial$}\lrcorner}\frac{1}{2}\left\langle F_{1}%
\theta^{\mathbf{a}}\tilde{F}_{2}+F_{2}\theta^{\mathbf{a}}\tilde{F}%
_{1}\right\rangle _{1}=0. \label{em16}%
\end{equation}

We just proved that indeed, $\delta\mathcal{K}^{\mathbf{a}}=0$, or what is the
same, that
\begin{equation}
d\star\mathcal{K}^{\mathbf{a}}=0,
\end{equation}
and since we are in Minkowski spacetime Poincar\'{e}'s lemma implies that the
$3$-form fields $\star\mathcal{K}^{\mathbf{a}}\in\sec%
{\textstyle\bigwedge\nolimits^{3}}
T^{\ast}M\hookrightarrow\sec\mathcal{C\ell(}M,\mathtt{\eta})$ must be exact,
i.e.,
\begin{equation}
\star\mathcal{K}^{\mathbf{a}}=-d\star\mathcal{E}^{\mathbf{a}}, \label{s1}%
\end{equation}
or
\begin{equation}
\delta\mathcal{E}^{\mathbf{a}}=-\mathcal{K}^{\mathbf{a}} \label{s1bis}%
\end{equation}

\section{The Energies and Momenta of Two Different Superposed Electromagnetic
Field Configurations are Additive.}

In this section the standard cylinder of Minkowski spacetime and its boundary
submanifolds (see Figure ) described in the Appendix will be used. We start
our enterprise by recalling that since $\star\mathcal{T}_{\mathbf{\ }%
}^{\mathbf{a}}=\star\mathcal{T}_{1\mathbf{\ \ }}^{\mathbf{a}}+\star
\mathcal{T}_{2\mathbf{\ \ }}^{\mathbf{a}}+\star\mathcal{K}^{\mathbf{a}}$ and
$d\star\mathcal{T}^{\mathbf{a}}=0$,$\ \ d\star\mathcal{T}_{1}^{\mathbf{a}}=0$,
$d\star\mathcal{T}_{2}^{\mathbf{a}}=0$ and $d\star\mathcal{K}^{\mathbf{a}}=0$
we can use Stokes theorem to write
\begin{align}
0  &  =%
{\displaystyle\int\nolimits_{N}}
d\star\mathcal{K}_{\mathbf{\ }}^{\mathbf{a}}=-%
{\displaystyle\int\nolimits_{B_{1}^{\prime}}}
\star\mathcal{K}_{\mathbf{\ }}^{\mathbf{a}}+%
{\displaystyle\int\nolimits_{B_{2}^{\prime}}}
\star\mathcal{K}_{\mathbf{\ }}^{\mathbf{a}}+%
{\displaystyle\int\nolimits_{B_{3}^{\prime}}}
\star\mathcal{K}_{\mathbf{\ }}^{\mathbf{a}}\nonumber\\
&  =-%
{\displaystyle\int\nolimits_{B_{1}}}
\star\mathcal{K}_{\mathbf{\ }}^{\mathbf{a}}+%
{\displaystyle\int\nolimits_{B_{2}}}
\star\mathcal{K}_{\mathbf{\ }}^{\mathbf{a}}+%
{\displaystyle\int\nolimits_{B_{3}^{\prime}}}
\star\mathcal{K}_{\mathbf{\ }}^{\mathbf{a}}, \label{N4}%
\end{align}
from where it follows that
\begin{equation}%
{\displaystyle\int\nolimits_{B_{1}}}
\star\mathcal{K}_{\mathbf{\ }}^{\mathbf{a}}=%
{\displaystyle\int\nolimits_{B_{2}}}
\star\mathcal{K}_{\mathbf{\ }}^{\mathbf{a}}+%
{\displaystyle\int\nolimits_{B_{3}^{\prime}}}
\star\mathcal{K}_{\mathbf{\ }}^{\mathbf{a}}. \label{N5}%
\end{equation}

Now, if we take into account Eq.(\ref{em11}) which defines $\star
\mathcal{K}_{\mathbf{\ }}^{\mathbf{a}}$ we see immediately that $%
{\displaystyle\int\nolimits_{B_{3}^{\prime}}}
\star\mathcal{K}_{\mathbf{\ }}^{\mathbf{a}}=0$ because $\left.  F_{1}%
\right\vert _{B_{3}^{\prime}}=0$ and $\left.  F_{1}\right\vert _{B_{3}%
^{\prime}}=0$. \ Also, since (obviously) $\left.  \star\mathcal{K}%
^{\mathbf{a}}\right\vert _{B_{1}}=0$ it follows that $%
{\displaystyle\int\nolimits_{B_{1}}}
\star\mathcal{K}_{\mathbf{\ }}^{\mathbf{a}}=0$ and we get that
\begin{equation}%
{\displaystyle\int\nolimits_{B_{2}}}
\star\mathcal{K}_{\mathbf{\ }}^{\mathbf{a}}=0, \label{N5bis}%
\end{equation}
even if \ $\left.  \star\mathcal{K}^{\mathbf{a}}\right\vert _{B_{2}}\neq0$.

Using these results we can calculate the total energy of $F=F_{1}+F_{2}$
containing in $B_{2}$. We have, taking into account Eq.(\ref{em10}) and
Eq.(\ref{N5bis}) that
\begin{align*}%
{\displaystyle\int\nolimits_{B_{2}}}
\star\mathcal{T}_{\mathbf{\ }}^{\mathbf{a}}  &  =%
{\displaystyle\int\nolimits_{B_{2}}}
\star(\mathcal{T}_{1\mathbf{\ }}^{\mathbf{a}}+\mathcal{T}_{2\mathbf{\ }%
}^{\mathbf{a}})+%
{\displaystyle\int\nolimits_{B_{2}}}
\star\mathcal{K}_{\mathbf{\ }}^{\mathbf{a}}\\
&  =%
{\displaystyle\int\nolimits_{B_{2}}}
\star(\mathcal{T}_{1\mathbf{\ }}^{\mathbf{a}}+\mathcal{T}_{2\mathbf{\ }%
}^{\mathbf{a}}),
\end{align*}
i.e.,
\begin{equation}
P_{F}^{\mathbf{a}}=P_{F_{1}}^{\mathbf{a}}+P_{F_{2}}^{\mathbf{a}},
\end{equation}
as we wanted to prove.

\section{Conclusions}

In this paper we proved that the energy and momenta of two free
electromagnetic field configurations (which satisfy at any instant of time a
free Maxwell equation and for each finite instant of time have compact support
in $\mathbb{R}^{3}$) of\textit{ finite energy }are additive and thus there is
no incompatibility between the principle of superposition of energy and the
principle of energy-momentum conservations as suggested by some authors,
quoted in Section 1\footnote{We observe that the same problem occur for all
linear field theories, and indeed in reference (\cite{levine,matews,rowpask})
we have some discussion of the problem for sound (and other elastic) waves.}.
We emphasize that our proof is made relatively simple due to the amazing power
of the Clifford bundle formalism, and indeed we do not see how to do the
calculations using the old Heaviside-Gibbs vector calculus, or even only the
Cartan calculus of differential forms, since our proof depends deeply on the
noticeable formula for the energy-momentum densities, namely $\star
\mathcal{T}^{\mathbf{a}}=\frac{1}{2}F\theta^{\mathbf{a}}\tilde{F}$ which is
valid for any electromagnetic field configuration $F$ satisfying Maxwell
equation ${\mbox{\boldmath$\partial$}}F={0}$. We emphasize also that we have
found there exist closed $2$-forms $\star\mathcal{W}^{\mathbf{a}}\in
\sec\bigwedge\nolimits^{2}T^{\ast}M\hookrightarrow\sec\mathcal{C\ell
(}M,\mathtt{\eta})$ satisfying generalized Maxwell equations
${\mbox{\boldmath$\partial$}}\mathcal{W}^{\mathbf{a}}=\mathcal{T}^{\mathbf{a}%
}+\star M_{\mathbf{a}}$. This result that will be explored in future publications.

\appendix

\section{Clifford Bundles}

Let $(M,\mathbf{\eta},D,\tau_{\mathtt{\eta}},\uparrow)$ be Minkowski
spacetime. $(M,\mathbf{\eta})$ is a four dimensional space oriented (by the
volume form $\tau_{\mathtt{\eta}})$ and time oriented (by the equivalence
relation $\uparrow$, see \cite{rodcap}) Lorentzian manifold, with
$M\simeq\mathbb{R}^{4}$ and $\mathbf{\eta}\in\sec T_{2}^{0}M$ is a Lorentzian
metric of signature $(1,3)$. $T^{\ast}M$ [$TM$] is the cotangent [tangent]
bundle. $T^{\ast}M=\cup_{x\in M}T_{x}^{\ast}M$, $TM=\cup_{x\in M}T_{x}M$, and
$T_{x}M\simeq T_{x}^{\ast}M\simeq\mathbb{R}^{1,3}$, where $\mathbb{R}^{1,3}$
is the Minkowski vector space~. $D$ is the Levi-Civita connection of
$\mathbf{\eta}$, $i.e$\textit{.\/}, $D\mathbf{\eta}=0$, $\mathbf{R}(D)=0$.
Also $\mathbf{T}(D)=0$, $\mathbf{R}$ and $\mathbf{T}$ being respectively the
torsion and curvature tensors. Let $\mathtt{\eta}\in\sec T_{0}^{2}M$ be the
metric on the cotangent bundle associated to $\mathbf{\eta}\in\sec T_{2}^{0}%
M$. The Clifford bundle of differential forms $\mathcal{C}\!\ell
(M,\mathtt{\eta})$ is the bundle of algebras, i.e., $\mathcal{C}%
\!\ell(M,\mathtt{\eta})=\cup_{x\in M}\mathcal{C}\!\ell(T_{x}^{\ast}M)$, where
$\forall x\in M$, $\mathcal{C}\!\ell(T_{x}^{\ast}M)=\mathbb{R}_{1,3}$, the so
called \emph{spacetime} \emph{algebra}. Recall also that $\mathcal{C}%
\!\ell(M,\mathtt{\eta})$ is a vector bundle associated to the
\emph{\ orthonormal frame bundle \ }$\mathbf{P}_{\mathrm{SO}_{(1,3)}^{e}}(M)$,
i.e., $\mathcal{C}\!\ell(M,\mathtt{\eta})$ $=P_{SO_{+(1,3)}}(M)\times
_{ad}\mathbb{R}_{1,3}$ (see details in, e.g., \cite{wal,walmosna,rodcap}). For
any $x\in M$, $\mathcal{C}\!\ell(T_{x}^{\ast}M)$ is a linear space over the
real field $\mathbb{R}$. Moreover, $\mathcal{C}\!\ell(T_{x}^{\ast}M)$ is
isomorphic to the Cartan algebra $\bigwedge T_{x}^{\ast}M$ of the cotangent
space and $\bigwedge T_{x}^{\ast}M=\sum_{k=0}^{4}\bigwedge{}^{k}T_{x}^{\ast}%
M$, where $\bigwedge^{k}T_{x}^{\ast}M$ is the $\binom{4}{k}$-dimensional space
of $k$-forms. Then, sections of $\mathcal{C}\!\ell(M,\mathtt{\eta})$ can be
represented as a sum of non homogeneous differential forms. Let $\{x^{\mu}\}$
be coordinates in Einstein-Lorentz-Poincar\'{e} gauge for $M$ and let
$\{e_{\mu}=\partial/\partial x^{\mu}\}\in\sec FM$ (the frame bundle) be an
orthonormal basis for $TM$, i.e., $\mathbf{\eta}(e_{\mu},e_{\nu})=\eta_{\mu
\nu}=\mathrm{diag}(1,-1,-1,-1)$, Let $\gamma^{\nu}=dx^{\nu}\in\sec
\bigwedge^{1}T^{\ast}M\hookrightarrow\sec\mathcal{C}\!\ell(M,\mathtt{\eta})$
($\nu=0,1,2,3$) such that the set $\{\gamma^{\nu}\}$ is the dual basis of
$\{e_{\mu}\},$ and of course, \ $\mathtt{\eta}(\gamma^{\mu},\gamma^{\nu}%
)=\eta^{\mu\nu}=\mathrm{diag}(1,-1,-1,-1)$. \ We introduce moreover the
notations $\theta^{\mathbf{a}}=\delta_{\mu}^{\mathbf{a}}dx^{\mu}$ and
$\mathbf{e}_{\mathbf{a}}=\delta_{\mathbf{a}}^{\mu}\frac{\partial}{\partial
x^{\mu}}$. We say that $\{\mathbf{e}_{\mathbf{a}}\}$ is a section of the
orthonormal frame bundle $\mathbf{P}_{\mathrm{SO}_{(1,3)}^{e}}(M)$ and its
dual basis $\{\theta^{\mathbf{a}}\}$ is a section of the orthonormal coframe
bundle (denoted $P_{\mathrm{SO}_{(1,3)}^{e}}(M)$).

\subsection{Clifford Product}

The fundamental \emph{Clifford product} (in what follows to be denoted by
juxtaposition of symbols) is generated by $\theta^{\mathbf{a}}\theta
^{\mathbf{a}}+\theta^{\mathbf{b}}\theta^{\mathbf{a}}=2\eta^{\mathbf{ab}}$ and
if $\mathcal{C}\in\mathcal{C}\!\ell(M,\mathtt{\eta})$ we have%

\begin{equation}
\mathcal{C}=s+v_{\mathbf{a}}\theta^{\mathbf{a}}+\frac{1}{2!}b_{\mathbf{ab}%
}\theta^{\mathbf{a}}\theta^{\mathbf{b}}+\frac{1}{3!}a_{\mathbf{abc}}%
\theta^{\mathbf{a}}\theta^{\mathbf{b}}\theta^{\mathbf{c}}+p\theta^{5}\;,
\label{3}%
\end{equation}
where $\tau_{\mathtt{\eta}}:=\theta^{5}=\theta^{0}\theta^{1}\theta^{2}%
\theta^{3}=dx^{0}dx^{1}dx^{2}dx^{3}$ is the volume element and $s$,
$v_{\mathbf{a}}$, $b_{\mathbf{ab}}$, $a_{\mathbf{abc}}$, $p\in\sec
\bigwedge^{0}T^{\ast}M\hookrightarrow\sec\mathcal{C}\!\ell(M,\mathtt{\eta})$.

Let $\mathcal{A}_{r},\in\sec\bigwedge^{r}T^{\ast}M\hookrightarrow
\sec\mathcal{C}\!\ell(M,\mathtt{\eta}),\mathcal{B}_{s}\in\sec\bigwedge
^{s}T^{\ast}M\hookrightarrow\sec\mathcal{C}\!\ell(M,\mathtt{\eta})$. For
$r=s=1$, we define the \emph{scalar product} as follows:

For $a,b\in\sec\bigwedge^{1}T^{\ast}M\hookrightarrow\sec\mathcal{C}%
\!\ell(M,\mathtt{\eta}),$%
\begin{equation}
a\cdot b=\frac{1}{2}(ab+ba)=\mathtt{\eta}(a,b). \label{4}%
\end{equation}
We define also the \emph{exterior product} ($\forall r,s=0,1,2,3)$ by
\begin{align}
\mathcal{A}_{r}\wedge\mathcal{B}_{s}  &  =\langle\mathcal{A}_{r}%
\mathcal{B}_{s}\rangle_{r+s},\nonumber\\
\mathcal{A}_{r}\wedge\mathcal{B}_{s}  &  =(-1)^{rs}\mathcal{B}_{s}%
\wedge\mathcal{A}_{r}, \label{5}%
\end{align}
where $\langle\rangle_{k}$ is the component in $\bigwedge^{k}T^{\ast}M$
\ (projection) of the Clifford field. The exterior product is extended by
linearity to all sections of $\mathcal{C}\!\ell(M,\mathtt{\eta})$..

For $\mathcal{A}_{r}=a_{1}\wedge...\wedge a_{r},$ $\mathcal{B}_{r}=b_{1}%
\wedge...\wedge b_{r}$, the scalar product is defined here as follows,
\begin{align}
\mathcal{A}_{r}\cdot\mathcal{B}_{r}  &  =(a_{1}\wedge...\wedge a_{r}%
)\cdot(b_{1}\wedge...\wedge b_{r})\nonumber\\
&  =\left\vert
\begin{array}
[c]{lll}%
a_{1}\cdot b_{1} & .... & a_{1}\cdot b_{r}\\
.......... & .... & ..........\\
a_{r}\cdot b_{1} & .... & a_{r}\cdot b_{r}%
\end{array}
\right\vert . \label{6}%
\end{align}

We agree that if $r=s=0$, the scalar product is simple the ordinary product in
the real field.

Also, if $r\neq s$, then $\mathcal{A}_{r}\cdot\mathcal{B}_{s}=0$. Finally, the
scalar product is extended by linearity for all sections of $\mathcal{C}%
\!\ell(M,\mathtt{\eta})$.

For $r\leq s,$ $\mathcal{A}_{r}=a_{1}\wedge...\wedge a_{r},$ $\mathcal{B}%
_{s}=b_{1}\wedge...\wedge b_{s\text{ }}$we define the left contraction by
\begin{equation}
\lrcorner:(\mathcal{A}_{r},\mathcal{B}_{s})\mapsto\mathcal{A}_{r}%
\lrcorner\mathcal{B}_{s}=%
{\displaystyle\sum\limits_{i_{1}\,<...\,<i_{r}}}
\epsilon^{i_{1}....i_{s}}(a_{1}\wedge...\wedge a_{r})\cdot(b_{_{i_{1}}}%
\wedge...\wedge b_{i_{r}})^{\sim}b_{i_{r}+1}\wedge...\wedge b_{i_{s}}
\label{7}%
\end{equation}
where $\sim$ is the reverse mapping (\emph{reversion}) defined by
\begin{equation}
\sim:\sec\bigwedge^{p}T^{\ast}M\ni a_{1}\wedge...\wedge a_{p}\mapsto
a_{p}\wedge...\wedge a_{1} \label{8}%
\end{equation}
and extended by linearity to all sections of $\mathcal{C}\!\ell(M,\mathtt{\eta
})$. We agree that for $\alpha,\beta\in\sec\bigwedge^{0}T^{\ast}M$ the
contraction is the ordinary (pointwise) product in the real field and that if
$\alpha\in\sec\bigwedge^{0}T^{\ast}M\hookrightarrow\mathcal{C}\!\ell
(M,\mathtt{\eta})$, $\mathcal{A}_{r},\in\sec\bigwedge^{r}T^{\ast
}M\hookrightarrow\mathcal{C}\!\ell(M,\mathtt{\eta})$, $\mathcal{B}_{s}\in
\sec\bigwedge^{s}T^{\ast}M$ $\hookrightarrow\mathcal{C}\!\ell(M,\mathtt{\eta
})$\ then $(\alpha\mathcal{A}_{r})\lrcorner\mathcal{B}_{s}=\mathcal{A}%
_{r}\lrcorner(\alpha\mathcal{B}_{s})$. Left contraction is extended by
linearity to all pairs of elements of sections of $\mathcal{C}\!\ell
(M,\mathtt{\eta})$, i.e., for $\mathcal{A},\mathcal{B}\in\sec\mathcal{C}%
\!\ell(M,\mathtt{\eta})$%

\begin{equation}
\mathcal{A}\lrcorner\mathcal{B}=\sum_{r,s}\langle\mathcal{A}\rangle
_{r}\lrcorner\langle\mathcal{B}\rangle_{s},\text{ }r\leq s. \label{9}%
\end{equation}

It is also necessary to introduce the operator of \emph{right contraction}
denoted by $\llcorner$. The definition is obtained from the one presenting the
left contraction with the imposition that $r\geq s$ and taking into account
that now if $\mathcal{A}_{r}\in\sec\bigwedge^{r}T^{\ast}M\hookrightarrow
\mathcal{C}\!\ell(M,\mathtt{\eta})$, $\mathcal{B}_{s}\in\sec\bigwedge
^{s}T^{\ast}M\hookrightarrow\mathcal{C}\!\ell(M,\mathtt{\eta})\ $then
$\mathcal{A}_{r}\llcorner(\alpha\mathcal{B}_{s})=(\alpha\mathcal{A}%
_{r})\llcorner\mathcal{B}_{s}$.

The main formulas used in the present paper can be obtained (details may be
found in \cite{rodcap}) from the following ones (where $a\in\sec\bigwedge
^{1}T^{\ast}M\hookrightarrow\sec\mathcal{C}\!\ell(M,\mathtt{\eta})$):
\begin{align}
a\mathcal{B}_{s} &  =a\lrcorner\mathcal{B}_{s}+a\wedge\mathcal{B}%
_{s},\mathcal{B}_{s}a=\mathcal{B}_{s}\llcorner a+\mathcal{B}_{s}\wedge
a,\nonumber\\
a\lrcorner\mathcal{B}_{s} &  =\frac{1}{2}(a\mathcal{B}_{s}-(-1)^{s}%
\mathcal{B}_{s}a),\nonumber\\
\mathcal{A}_{r}\lrcorner\mathcal{B}_{s} &  =(-1)^{r(s-r)}\mathcal{B}%
_{s}\llcorner\mathcal{A}_{r},\nonumber\\
a\wedge\mathcal{B}_{s} &  =\frac{1}{2}(a\mathcal{B}_{s}+(-1)^{s}%
\mathcal{B}_{s}a),\nonumber\\
\mathcal{A}_{r}\mathcal{B}_{s} &  =\langle\mathcal{A}_{r}\mathcal{B}%
_{s}\rangle_{|r-s|}+\langle\mathcal{A}_{r}\mathcal{B}_{s}\rangle
_{|r-s|+2}+...+\langle\mathcal{A}_{r}\mathcal{B}_{s}\rangle_{|r+s|}\nonumber\\
&  =\sum\limits_{k=0}^{m}\langle\mathcal{A}_{r}\mathcal{B}_{s}\rangle
_{|r-s|+2k}\nonumber\\
\mathcal{A}_{r}\cdot\mathcal{B}_{r} &  =\mathcal{B}_{r}\cdot\mathcal{A}%
_{r}=\widetilde{\mathcal{A}}_{r}\text{ }\lrcorner\mathcal{B}_{r}%
=\mathcal{A}_{r}\llcorner\widetilde{\mathcal{B}}_{r}=\langle\widetilde
{\mathcal{A}}_{r}\mathcal{B}_{r}\rangle_{0}=\langle\mathcal{A}_{r}%
\widetilde{\mathcal{B}}_{r}\rangle_{0},
\end{align}%
\begin{align}
\langle\mathcal{AB}\rangle_{r}  & =(-1)^{r(r-1)/2}\langle\widetilde
{\mathcal{B}}\widetilde{\mathcal{A}}\rangle_{r},\nonumber\\
\langle\mathcal{A}_{r}\mathcal{B}_{s}\rangle_{r}  & =\langle\widetilde
{\mathcal{B}}_{s}A_{r}\rangle_{r}=(-1)^{s(s-1)/2}\langle\mathcal{B}%
_{s}\mathcal{A}_{r}\rangle_{r},\nonumber\\
\langle\mathcal{A}_{r}\mathcal{B}_{s}\mathcal{C}_{t}\rangle_{q}  &
=(-1)^{\varepsilon}\langle\mathcal{C}_{t}\mathcal{B}_{s}\mathcal{A}_{r}%
\rangle_{q},\nonumber\\
\varepsilon & =\frac{1}{2}(q^{2}+r^{2}+s^{2}+t^{2}-q-r-s-t)\label{identities}%
\end{align}

\subsection{Hodge Star Operator}

Let $\star$ be the Hodge star operator, i.e., the mapping
\[
\star:\bigwedge^{k}T^{\ast}M\rightarrow\bigwedge^{4-k}T^{\ast}M,\text{
}\mathcal{A}_{k}\mapsto\star\mathcal{A}_{k}%
\]
where for $\mathcal{A}_{k}\in\sec\bigwedge^{k}T^{\ast}M\hookrightarrow
\mathcal{C}\!\ell(M,\mathtt{\eta})$%
\begin{equation}
\lbrack\mathcal{B}_{k}\cdot\mathcal{A}_{k}]\tau_{\mathtt{\eta}}=\mathcal{B}%
_{k}\wedge\star\mathcal{A}_{k},\text{ }\forall\mathcal{B}_{k}\in\sec
\bigwedge\nolimits^{k}T^{\ast}M\hookrightarrow\sec\mathcal{C}\!\ell
(M,\mathtt{\eta}). \label{11a}%
\end{equation}
$\tau_{\mathtt{\eta}}=\theta^{5}\in\bigwedge^{4}T^{\ast}M$ is a
\emph{standard} volume element. Then we can verify that
\begin{equation}
\star\mathcal{A}_{k}=\widetilde{\mathcal{A}}_{k}\tau_{\mathtt{\eta}%
}=\widetilde{\mathcal{A}}_{k}\theta^{5}. \label{11b}%
\end{equation}

\subsection{Dirac Operator}

Let $d$ and $\delta$ be respectively the differential and Hodge codifferential
operators acting on sections of $\sec\bigwedge\nolimits^{k}T^{\ast
}M\hookrightarrow\sec\mathcal{C}\!\ell(M,\mathtt{\eta})$. If $\mathcal{A}%
_{p}\in\sec\bigwedge^{p}T^{\ast}M\hookrightarrow\sec\mathcal{C}\!\ell
(M,\mathtt{\eta})$, then $\delta\mathcal{A}_{p}=(-1)^{p}\star^{-1}%
d\star\mathcal{A}_{p}$, with $\star^{-1}\star=\mathrm{identity}$.

The Dirac operator acting on sections of $\mathcal{C}\!\ell(M,g)$ is the
invariant first order differential operator
\begin{equation}
{\mbox{\boldmath$\partial$}}=\theta^{\mathbf{a}}D_{\mathbf{e}_{\mathbf{a}}}.
\label{12}%
\end{equation}

\subsubsection{$D_{\mathbf{e}_{\mathbf{a}}}\mathcal{A}$}

The \textit{reciprocal }basis of $\{\theta^{\mathbf{b}}\}$ is denoted
$\{\theta_{\mathbf{a}}\}$ and we have $\theta_{\mathbf{a}}\cdot\theta
_{\mathbf{b}}=\eta_{\mathbf{ab}}$ ($\eta_{\mathbf{ab}}=\mathrm{diag}%
(1,-1,-1,-1)$). Also,
\begin{equation}
D_{\mathbf{e}_{\mathbf{a}}}\theta^{\mathbf{b}}=-\omega_{\mathbf{ac}%
}^{\mathbf{b}}\theta^{\mathbf{c}}=-\omega_{\mathbf{a}}^{\mathbf{bc}}%
\theta_{\mathbf{c}}, \label{12n}%
\end{equation}
with $\omega_{\mathbf{a}}^{\mathbf{bc}}=-\omega_{\mathbf{a}}^{\mathbf{cb}}$,
and $\omega_{\mathbf{a}}^{\mathbf{bc}}=\eta^{\mathbf{bk}}\omega_{\mathbf{kal}%
}\eta^{\mathbf{cl}},$ $\omega_{\mathbf{abc}}=\eta_{\mathbf{ad}}\omega
_{\mathbf{bc}}^{\mathbf{d}}=-\omega_{\mathbf{cba}}$. Defining
\begin{equation}
\mathbf{\omega}_{\mathbf{a}}=\frac{1}{2}\omega_{\mathbf{a}}^{\mathbf{bc}%
}\theta_{\mathbf{b}}\wedge\theta_{\mathbf{c}}\in\sec%
{\displaystyle\bigwedge\nolimits^{2}}
T^{\ast}M\hookrightarrow\sec\mathcal{C}\!\ell(M,\mathtt{\eta}), \label{12nn}%
\end{equation}
we have (by linearity) that for any $\mathcal{A}\in\sec\bigwedge T^{\ast
}M\hookrightarrow\sec\mathcal{C}\!\ell(M,\mathtt{\eta})$
\begin{equation}
D_{\mathbf{e}_{\mathbf{a}}}\mathcal{A}=\partial_{\mathbf{e}_{\mathbf{a}}%
}\mathcal{A}+\frac{1}{2}[\mathbf{\omega}_{\mathbf{a}},\mathcal{A}],
\label{12nnn}%
\end{equation}
where $\partial_{\mathbf{e}_{\mathbf{a}}}$ is the Pfaff
derivative\footnote{E.g., if $A=\frac{1}{p!}A_{\mathbf{i}_{1}...\mathbf{i}%
_{p}}\theta^{_{\mathbf{i}_{1}}}...\theta^{_{.\mathbf{i}_{p}\text{ }}}$then
$\partial_{\mathbf{e}_{\mathbf{a}}}A=\frac{1}{p!}[\mathbf{e}_{\mathbf{a}%
}(A_{\mathbf{i}_{1}...\mathbf{i}_{p}})]\theta^{_{\mathbf{i}_{1}}}%
...\theta^{_{.\mathbf{i}_{p}\text{ }}}$.}

\subsubsection{${\mbox{\boldmath$\partial$}}=$ $d-{\delta}$}

Using Eq.(\ref{12nnn}) we can easily show the very important result:%

\begin{align}
{\mbox{\boldmath$\partial$}}\mathcal{A}  &  ={\mbox{\boldmath$\partial$}}%
\wedge\mathcal{A}+\,{\mbox{\boldmath$\partial$}}\lrcorner\mathcal{A}%
=d\mathcal{A}-\delta\mathcal{A},\nonumber\\
{\mbox{\boldmath$\partial$}}\wedge\mathcal{A}  &  =d\mathcal{A},\hspace
{0.1in}\,{\mbox{\boldmath$\partial$}}\lrcorner\mathcal{A}=-\delta\mathcal{A}.
\label{13}%
\end{align}

\section{Maxwell Equation}

Eq.(\ref{13}) permit us to write the Maxwell equations
\begin{equation}
dF=0,\text{ }\delta F=-J \label{14}%
\end{equation}
for $F\in\sec%
{\displaystyle\bigwedge\nolimits^{2}}
T^{\ast}M\hookrightarrow\sec\mathcal{C}\!\ell(M,\mathtt{\eta})$\ as a
\textit{single} equation (Maxwell equation),%
\begin{equation}
{\mbox{\boldmath$\partial$}}F=J. \label{Maxwell}%
\end{equation}

\subsection{The Noticeable Riesz Formula $\mathcal{T}_{\mathbf{a}}=\frac{1}%
{2}F\theta_{\mathbf{a}}\tilde{F}$}

We now prove that the energy-momentum densities $\star\mathcal{T}_{\mathbf{a}%
}$ of the Maxwell field can be written in the Clifford bundle formalism
as\footnote{The formula $\mathcal{T}_{\mathbf{a}}=\frac{1}{2}F\theta
_{\mathbf{a}}\tilde{F}$ has been first obtained (but, not using the algebraic
derivatives of the Lagrangian density) by M. Riesz in 1947 \cite{riesz} and it
has been rediscovered by Hestenes in 1996 \cite{hestenes} (which also does not
use the algebraic derivatives of the Lagrangian density). Algebraic
derivatives of \textit{homogenous} form fields has been described, e.g., in
Thirring's book \cite{thirring}.}:
\begin{equation}
\star\mathcal{T}_{\mathbf{a}}=\frac{1}{2}\star(F\theta_{\mathbf{a}}\tilde
{F})\in\sec\bigwedge\nolimits^{3}T^{\ast}M\hookrightarrow\sec\mathcal{C\ell
(}M,\mathtt{\eta)}. \label{em1}%
\end{equation}

To derive Eq.(\ref{em1}) we start from the Maxwell Lagrangian
\begin{equation}
\mathcal{L}_{m}=\frac{1}{2}F\wedge\star F, \label{em2}%
\end{equation}
where $F=\frac{1}{2}F_{\mathbf{ab}}\theta^{\mathbf{a}}\wedge\theta
^{\mathbf{b}}:=\frac{1}{2}F_{\mathbf{ab}}\theta^{\mathbf{ab}}\in\sec%
{\displaystyle\bigwedge\nolimits^{2}}
TM\hookrightarrow\sec\mathcal{C\ell(}M,\mathtt{\eta)}$ is the electromagnetic
field. Now, denoting by $%
\mbox{\boldmath{$\delta$}}%
$ the variational symbol\footnote{Please, do not confuse the variational
symbol $%
\mbox{\boldmath{$\delta$}}%
$ with the symbol $\delta$ of the Hodge coderiviative.} we can easily verify
that
\[%
\mbox{\boldmath{$\delta$}}%
\star\theta^{\mathbf{ab}}=%
\mbox{\boldmath{$\delta$}}%
\theta^{\mathbf{c}}\wedge\lbrack\theta_{\mathbf{c}}\lrcorner\star
\theta^{\mathbf{ab}}].
\]
Moreover, in general $%
\mbox{\boldmath{$\delta$}}%
$ and $\star$ do not commute. Indeed,\ for any \ $\mathcal{A}_{p}\in\sec%
{\displaystyle\bigwedge\nolimits^{p}}
T^{\ast}M\hookrightarrow\sec\mathcal{C\ell(}M,\mathtt{\eta)}$ we have
\begin{align}
\lbrack%
\mbox{\boldmath{$\delta$}}%
,\star]\mathcal{A}_{p}  &  =%
\mbox{\boldmath{$\delta$}}%
\star\mathcal{A}_{p}-\star%
\mbox{\boldmath{$\delta$}}%
\mathcal{A}_{p}\label{7.ex00}\\
&  =%
\mbox{\boldmath{$\delta$}}%
\theta^{\mathbf{a}}\wedge\left(  \theta_{\mathbf{a}}\lrcorner\star
\mathcal{A}_{p}\right)  -\star\left[
\mbox{\boldmath{$\delta$}}%
\theta^{\mathbf{a}}\wedge\left(  \theta_{\mathbf{a}}\lrcorner\mathcal{A}%
_{p}\right)  \right]  .\nonumber
\end{align}
Multiplying both members of Eq.(\ref{7.ex00}) with $\mathcal{A}_{p}=F$ on the
right by $F\wedge$ we get%
\[
F\wedge%
\mbox{\boldmath{$\delta$}}%
\star F=F\wedge\star%
\mbox{\boldmath{$\delta$}}%
F+F\wedge\{%
\mbox{\boldmath{$\delta$}}%
\theta^{\mathbf{a}}\wedge(\theta_{\mathbf{a}}\lrcorner\star F)-\star\lbrack%
\mbox{\boldmath{$\delta$}}%
\theta^{\mathbf{a}}\wedge(\theta_{\mathbf{a}}\lrcorner F)]\}.
\]

Next we sum $%
\mbox{\boldmath{$\delta$}}%
F\wedge\star F$ to both members of the above equation obtaining%

\[%
\mbox{\boldmath{$\delta$}}%
\left(  F\wedge\star F\right)  =2%
\mbox{\boldmath{$\delta$}}%
F\wedge\star F+%
\mbox{\boldmath{$\delta$}}%
\theta^{\mathbf{a}}\wedge\lbrack F\wedge(\theta_{\mathbf{a}}\lrcorner\star
F)-(\theta_{\mathbf{a}}\lrcorner F)\wedge\star F].
\]

Then, it follows (see, \cite{rodcap,rodquinro} for details) that
if\footnote{$\pounds _{\xi}$ denotes the Lie derivative in the direction of
the vector field $\xi$.} $%
\mbox{\boldmath{$\delta$}}%
\theta^{\mathbf{a}}=-\pounds _{\xi}\theta^{\mathbf{a}}$, for some
diffeomorphism generated by the vector field $\xi$ that
\[
\star\mathcal{T}_{\mathbf{a}}=\frac{\partial\mathcal{L}_{m}}{\partial
\theta^{\mathbf{a}}}=\frac{1}{2}\left[  F\wedge(\theta_{\mathbf{a}}%
\lrcorner\star F)-(\theta_{\mathbf{a}}\lrcorner F)\wedge\star F\right]  .
\]
Now,
\[
(\theta_{\mathbf{a}}\lrcorner F)\wedge\star F=-\star\lbrack(\theta
_{\mathbf{a}}\lrcorner F)\lrcorner F]=-[(\theta_{\mathbf{a}}\lrcorner
F)\lrcorner F]\tau_{\mathtt{\mathbf{\eta}}}%
\]
and using also the identity \cite{rodcap}
\[
(\theta_{\mathbf{a}}\lrcorner F)\wedge\star F=\theta_{\mathbf{a}}(F\cdot
F)\tau_{\mathtt{\eta}}-F\wedge(\theta_{\mathbf{a}}\lrcorner\star F).
\]
we get%

\begin{align*}
\frac{1}{2}\left[  F\wedge(\theta_{\mathbf{a}}\lrcorner\star F)-(\theta
_{\mathbf{a}}\lrcorner F)\wedge\star F\right]   &  =\frac{1}{2}\left\{
\theta_{\mathbf{a}}(F\cdot F)\tau_{\mathtt{\mathbf{\eta}}}-(\theta
_{\mathbf{a}}\lrcorner F)\wedge\star F-(\theta_{\mathbf{a}}\lrcorner
F)\wedge\star F\right\} \\
&  =\frac{1}{2}\left\{  \theta_{\mathbf{a}}(F\cdot F)\tau
_{\mathtt{\mathbf{\eta}}}-2(\theta_{\mathbf{a}}\lrcorner F)\wedge\star
F\right\} \\
&  =\frac{1}{2}\left\{  \theta_{\mathbf{a}}(F\cdot F)\tau
_{\mathtt{\mathbf{\eta}}}+2[(\theta_{\mathbf{a}}\lrcorner F)\lrcorner
F]\tau_{\mathtt{\mathbf{\eta}}}\right\} \\
&  =\star\left(  \frac{1}{2}\theta_{\mathbf{a}}(F\cdot F)+(\theta_{\mathbf{a}%
}\lrcorner F)\lrcorner F\right)  =\frac{1}{2}\star(F\theta_{\mathbf{a}}%
\tilde{F}),
\end{align*}
where in writing the last line we used the identity
\begin{equation}
\frac{1}{2}Fn\tilde{F}=(n\lrcorner F)\lrcorner F+\frac{1}{2}n(F\cdot F),
\label{6.66a}%
\end{equation}
whose proof is as follows:
\begin{align*}
(n\lrcorner F)\lrcorner F+\frac{1}{2}n(F\cdot F)  &  =\frac{1}{2}\left[
(n\lrcorner F)F-F(n\lrcorner F)\right]  +\frac{1}{2}n(F\cdot F)\\
&  =\frac{1}{4}\left[  nFF-FnF-FnF+FFn\right]  +\frac{1}{2}n(F\cdot F)\\
&  =-\frac{1}{2}FnF+\frac{1}{4}\left[  -2n(F\cdot F)+n(F\wedge F)+(F\wedge
F)n\right]  +\frac{1}{2}n(F\cdot F)\\
&  =-\frac{1}{2}FnF+-\frac{1}{2}n(F\cdot F)+\frac{1}{2}n\wedge(F\wedge
F)+\frac{1}{2}n(F\cdot F)\\
&  =-\frac{1}{2}FnF=\frac{1}{2}Fn\tilde{F}.
\end{align*}
valid for any $n\in\sec%
{\displaystyle\bigwedge\nolimits^{1}}
T^{\ast}M\hookrightarrow\sec\mathcal{C\ell}(M,\mathtt{\eta})$ \ and \ $F$
$\in\sec%
{\displaystyle\bigwedge\nolimits^{2}}
T^{\ast}M\hookrightarrow\sec\mathcal{C\ell}(M,\mathtt{\eta})$.

For completeness and presentation of \ some more tricks of the trade we detail
the proof that $\mathcal{T}_{\mathbf{a}}\cdot\theta_{\mathbf{b}}%
=\mathcal{T}_{\mathbf{b}}\cdot\theta_{\mathbf{a}}.$
\begin{align*}
\mathcal{T}_{\mathbf{a}}\cdot\theta_{\mathbf{b}}  &  =-\frac{1}{2}\langle
F\theta_{\mathbf{a}}F\theta_{\mathbf{b}}\rangle_{0}=-\langle(F\llcorner
\theta_{\mathbf{a}})F\theta_{\mathbf{b}}\rangle_{0}-\frac{1}{2}\langle
(\theta_{\mathbf{a}}\lrcorner F\text{ }+\theta_{\mathbf{a}}\wedge F)\text{
}F\theta_{\mathbf{b}}\rangle_{0}\\
&  =-\langle(F\llcorner\theta_{\mathbf{a}})F\theta_{\mathbf{b}}\rangle
_{0}-\frac{1}{2}\langle(\theta_{\mathbf{a}}FF\theta_{\mathbf{b}}\rangle_{0}\\
&  =-\langle(F\llcorner\theta_{\mathbf{a}})(F\llcorner\theta_{\mathbf{b}%
})+(F\llcorner\theta_{\mathbf{a}})(F\wedge\theta_{\mathbf{b}})\rangle
_{0}+\frac{1}{2}\langle\theta_{\mathbf{a}}(F\cdot F)\theta^{\mathbf{b}}%
\rangle_{0}-\frac{1}{2}\langle\text{ }\theta_{\mathbf{a}}(F\wedge F)\text{
}\theta_{\mathbf{b}}\rangle_{0}\\
&  =-\langle(F\llcorner\theta_{\mathbf{a}})(F\llcorner\theta_{\mathbf{b}%
})\rangle_{0}+\frac{1}{2}\langle(F\cdot F)(\theta_{\mathbf{a}}\cdot
\theta_{\mathbf{b}})\rangle_{0}\\
&  =-(F\llcorner\theta_{\mathbf{b}})\cdot(F\llcorner\theta_{\mathbf{a}}%
)+\frac{1}{2}(F\cdot F)(\theta_{\mathbf{b}}\cdot\theta_{\mathbf{a}%
})=\mathcal{T}_{\mathbf{b}}\cdot\theta_{\mathbf{a}}.
\end{align*}

Note moreover that%
\begin{equation}
\mathcal{T}_{\mathbf{ab}}=\mathcal{T}_{\mathbf{a}}\cdot\theta_{\mathbf{b}%
}=-\eta^{\mathbf{cl}}F_{\mathbf{ac}}F_{\mathbf{bl}}+\frac{1}{4}F_{\mathbf{cd}%
}F^{\mathbf{cd}}\eta_{\mathbf{ab}},
\end{equation}
a well known result.

Of course, for the \textit{free} electromagnetic field we have that
$d\star\mathcal{T}^{\mathbf{a}}=0$, which is equivalent to $\delta
\mathcal{T}^{\mathbf{a}}=-{\mbox{\boldmath$\partial$}\lrcorner\mathcal{T}%
^{\mathbf{a}}=0}$. Indeed, observe that
\begin{align}
{\mbox{\boldmath$\partial$}\lrcorner\mathcal{T}^{\mathbf{a}}}%
{=\mbox{\boldmath$\partial$}\lrcorner}  &  \frac{1}{2}(F\theta^{\mathbf{a}%
}\tilde{F})\nonumber\\
&  =\frac{1}{2}\langle{\mbox{\boldmath$\partial$}}(F\theta^{\mathbf{a}}%
\tilde{F})\rangle_{0}\nonumber\\
&  =\frac{1}{2}\langle({\mbox{\boldmath$\partial$}}F)\theta^{\mathbf{a}}%
\tilde{F}+{\theta}^{\mathbf{b}}\left(  F\theta^{\mathbf{a}}D_{e_{\mathbf{b}}%
}\tilde{F}\right)  \rangle_{0}\nonumber\\
&  =\frac{1}{2}\langle{\theta}^{\mathbf{b}}\left(  F\theta^{\mathbf{a}%
}D_{e_{\mathbf{b}}}\tilde{F}\right)  \rangle_{0},
\end{align}
where we used that ${\mbox{\boldmath$\partial$}}F=0$. Now,%

\begin{equation}%
\begin{array}
[c]{ll}%
{\theta}^{\mathbf{b}}\left(  F\theta^{\mathbf{a}}D_{e_{\mathbf{b}}}\tilde
{F}\right)  & ={\theta}^{\mathbf{b}}\left\langle F\theta^{\mathbf{a}%
}D_{e_{\mathbf{b}}}\tilde{F}\right\rangle _{1}+{\theta}^{\mathbf{b}%
}\left\langle F\theta^{\mathbf{a}}D_{e_{\mathbf{b}}}\tilde{F}\right\rangle
_{3}\\
& ={\theta}^{\mathbf{b}}\lrcorner\left\langle F\theta^{\mathbf{a}%
}D_{e_{\mathbf{b}}}\tilde{F}\right\rangle _{1}+{\theta}^{\mathbf{b}}%
\wedge\left\langle F\theta^{\mathbf{a}}D_{e_{\mathbf{b}}}\tilde{F}%
\right\rangle _{1}\\
& +{\theta}^{\mathbf{b}}\lrcorner\left\langle F\theta^{\mathbf{a}%
}D_{e_{\mathbf{b}}}\tilde{F}\right\rangle _{3}+{\theta}^{\mathbf{b}}%
\wedge\left\langle F\theta^{\mathbf{a}}D_{e_{\mathbf{b}}}\tilde{F}%
\right\rangle _{3}.
\end{array}
\end{equation}
Then%
\begin{align*}
\langle{\theta}^{\mathbf{b}}\left(  F\theta^{\mathbf{a}}D_{e_{\mathbf{b}}%
}\tilde{F}\right)  \rangle_{0}  &  ={\theta}^{\mathbf{b}}\lrcorner\left\langle
F\theta^{\mathbf{a}}D_{e_{\mathbf{b}}}\tilde{F}\right\rangle _{1}=\left\langle
F\theta^{\mathbf{a}}D_{e_{\mathbf{b}}}\tilde{F}\right\rangle _{1}%
\llcorner{\theta}^{\mathbf{b}}\\
&  =\left\langle F\theta^{\mathbf{a}}D_{e_{\mathbf{b}}}\tilde{F}{\theta
}^{\mathbf{b}}\right\rangle _{0}\\
&  =\langle F\theta^{\mathbf{a}}\widetilde{({\mbox{\boldmath$\partial$}}%
F)}\rangle_{0}=0,
\end{align*}
where we used the symbol $\widetilde{({\mbox{\boldmath$\partial$}}%
F)}:=D_{e_{\mathbf{b}}}\tilde{F}{\theta}^{\mathbf{b}}$ and the fact that
\ $\widetilde{({\mbox{\boldmath$\partial$}}F)}=0$.

\subsection{ Enter New Maxwell Like Equations $d\star\mathcal{W}^{\mathbf{a}%
}=-\star\mathcal{T}^{\mathbf{a}}$, $d\mathcal{W}^{\mathbf{a}}=\star
M^{\mathbf{a}}$}

Let $\star\mathcal{T}^{\mathbf{a}}=\frac{1}{2}\star(F\theta^{\mathbf{a}}%
\tilde{F})\in\sec%
{\textstyle\bigwedge\nolimits^{3}}
T^{\ast}M\hookrightarrow\sec\mathcal{C\ell}(M,\mathtt{\eta})$ be the
energy-momentum densities of a free electromagnetic field configuration
$F\in\sec%
{\textstyle\bigwedge\nolimits^{2}}
T^{\ast}M\hookrightarrow\sec\mathcal{C\ell}(M,\mathtt{\eta})$
$({\mbox{\boldmath$\partial$}}F=0)$. As we already know, we have
\begin{equation}
-\delta\mathcal{T}^{\mathbf{a}}={\mbox{\boldmath$\partial$}\lrcorner
}\mathcal{T}^{\mathbf{a}}=0. \label{J1}%
\end{equation}
Eq.(\ref{J1}) is equivalent to $d\star\mathcal{T}^{\mathbf{a}}=0$ and since we
are in Minkowski spacetime there must exist $\mathcal{W}^{\mathbf{a}}\in\sec%
{\textstyle\bigwedge\nolimits^{2}}
T^{\ast}M\hookrightarrow\sec\mathcal{C\ell}(M,\mathtt{\eta})$ such that
\begin{equation}
-\mathcal{T}^{\mathbf{a}}=\delta\mathcal{W}^{\mathbf{a}}. \label{J2}%
\end{equation}

Of course, we must also have
\begin{equation}
d\mathcal{W}^{\mathbf{a}}=\star M^{\mathbf{a}},\label{J3}%
\end{equation}
for some $M^{\mathbf{a}}\in\sec%
{\textstyle\bigwedge\nolimits^{1}}
T^{\ast}M\hookrightarrow\sec\mathcal{C\ell}(M,\mathtt{\eta})$. Eqs. (\ref{J2})
and (\ref{J3}) may be writen as ${\mbox{\boldmath$\partial$}}\mathcal{W}%
^{\mathbf{a}}=\star M^{\mathbf{a}}+\mathcal{T}^{\mathbf{a}}$. In another
publication \cite{notte} we  determine the explicit form of the $\mathcal{W}%
^{\mathbf{a}}$ and the $M^{\mathbf{a}}.$

\section{Standard Cylinder $N$ in Minkowski Spacetime and its Boundary
Submanifolds}

\ Let $N$ be the standard cylinder (Figure 1 at the end of the paper)
\cite{sawu} in Minkowski spacetime described in the
Einstein-Lorentz-Poincar\'{e} coordinates $\{x^{\mu}\}$ naturally adapted to a
inertial frame $\mathbf{I}=\partial/\partial x^{0}$ by%
\begin{equation}
N=\left\{  (x^{0},x^{1},x^{2},x^{3})\}\text{ }|\text{ }%
{\displaystyle\sum\limits_{i=1}^{3}}
x^{i}x^{i}\leq r^{\prime}\text{, }0\leq x^{0}\leq\mathfrak{t}\right\}
\label{N1}%
\end{equation}

The boundary manifolds of $N$ are the following submanifolds of $M$,%
\begin{align}
B_{1}^{\prime}  &  =\left\{  (0,x^{1},x^{2},x^{3})\}\text{ }|\text{ }%
{\displaystyle\sum\limits_{i=1}^{3}}
x^{i}x^{i}<r^{\prime}\right\} \nonumber\\
B_{2}^{\prime}  &  =\left\{  (\mathfrak{t},x^{1},x^{2},x^{3})\}\text{ }|\text{
}%
{\displaystyle\sum\limits_{i=1}^{3}}
x^{i}x^{i}<r^{\prime}\right\} \nonumber\\
B_{3}^{\prime}  &  =\left\{  (x^{0},x^{1},x^{2},x^{3})\}\text{ }|\text{ }%
{\displaystyle\sum\limits_{i=1}^{3}}
x^{i}x^{i}=r^{\prime}\text{, }0<x^{0}<\mathfrak{t}\right\} \nonumber\\
C_{1}^{\prime}  &  =\left\{  (0,x^{1},x^{2},x^{3})\}\text{ }|\text{ }%
{\displaystyle\sum\limits_{i=1}^{3}}
x^{i}x^{i}=r^{\prime}\right\} \nonumber\\
C_{2}^{\prime}  &  =\left\{  (\mathfrak{t},x^{1},x^{2},x^{3})\}\text{ }|\text{
}%
{\displaystyle\sum\limits_{i=1}^{3}}
x^{i}x^{i}=r^{\prime}\right\}  , \label{N2}%
\end{align}
where $B_{3}^{\prime}$ is a timelike hypersurface and the other four are
spacelike hypersurfaces. We define also the manifolds $B_{1}\subset
B_{1}^{\prime}$ and $B_{2}\subset B_{2}^{\prime}$ \
\begin{align}
B_{1}  &  =\left\{  (0,x^{1},x^{2},x^{3})\}\text{ }|\text{ }%
{\displaystyle\sum\limits_{i=1}^{3}}
x^{i}x^{i}<r_{1}\text{, }r_{1}<<r^{\prime}\right\}  ,\nonumber\\
B_{2}  &  =\left\{  (\mathfrak{t},x^{1},x^{2},x^{3})\}\text{ }|\text{ }%
{\displaystyle\sum\limits_{i=1}^{3}}
x^{i}x^{i}<r_{2}\text{, }r_{2}<<r^{\prime}\right\}  , \label{N3}%
\end{align}
which contain respectively ( see Figure 1) the field configurations
$F(0,\mathbf{x})=F_{1}(0,\mathbf{x)+}F_{2}(0,\mathbf{x)}$ and $F(\mathfrak{t}%
,\mathbf{x})=F_{1}(\mathfrak{t},\mathbf{x)+}F_{1}(\mathfrak{t},\mathbf{x)}$.

We denote the interior of $N$ by $U^{\prime}$ and also introduce the
submanifold $U\subset U^{\prime}$ (Figure 1). Table 1 collects \cite{sawu} the
main features of the above submanifolds, necessary for the integrations
(appearing in Stokes theorem) performed in the main text
\[%
\begin{array}
[c]{ccccc}%
\text{{\small Submanifold}} & \text{{\small Topology}} &
\text{{\small Orientation}} & \text{{\small Closure}} &
\begin{array}
[c]{c}%
\text{{\small Causal}}\\
\text{{\small Character}}%
\end{array}
\\%
\begin{array}
[c]{c}%
U\\
U^{\prime}%
\end{array}
&
\begin{array}
[c]{c}%
\mathbb{R}^{4}\\
\mathbb{R}^{4}%
\end{array}
&
\begin{array}
[c]{c}%
\text{{\small from }}M\\
\text{{\small from }}M
\end{array}
&
\begin{array}
[c]{c}%
U^{-}=U%
{\displaystyle\bigcup\limits_{j=1}^{3}}
B_{j}%
{\displaystyle\bigcup\limits_{i=1}^{2}}
C_{i}\\
U^{\prime-}=U^{\prime}%
{\displaystyle\bigcup\limits_{j=1}^{3}}
B_{j}^{\prime}%
{\displaystyle\bigcup\limits_{i=1}^{2}}
C_{i}^{\prime}%
\end{array}
& \text{{\small timelike}}\\%
\begin{array}
[c]{c}%
B_{i}\text{ }(i=1,2)\\
B_{i}^{\prime}\text{ }(i=1,2)
\end{array}
&
\begin{array}
[c]{c}%
\mathbb{R}^{3}\\
\mathbb{R}^{3}%
\end{array}
&
\begin{array}
[c]{c}%
\text{{\small from }}U\\
\text{{\small from }}U^{\prime}%
\end{array}
&
\begin{array}
[c]{c}%
B_{i}^{-}=B_{i}%
{\displaystyle\bigcup}
C_{i}\\
B_{i}^{\prime-}=B_{i}^{\prime}%
{\displaystyle\bigcup}
C_{i}^{\prime}%
\end{array}
& \text{{\small spacelike}}\\%
\begin{array}
[c]{c}%
B_{3}\\
B_{3}^{\prime}%
\end{array}
&
\begin{array}
[c]{c}%
\mathbb{R\times}S^{2}\\
\mathbb{R\times}S^{2}%
\end{array}
&
\begin{array}
[c]{c}%
\text{{\small from }}U\\
\text{{\small from }}U^{\prime}%
\end{array}
&
\begin{array}
[c]{c}%
B_{3}^{-}=B_{3}%
{\displaystyle\bigcup}
C_{1}%
{\displaystyle\bigcup}
C_{2}\\
B_{3}^{\prime-}=B_{3}^{\prime}%
{\displaystyle\bigcup}
C_{1}^{\prime}%
{\displaystyle\bigcup}
C_{2}^{\prime}%
\end{array}
& \text{{\small timelike}}\\%
\begin{array}
[c]{c}%
C_{i}(i=1,2)\\
C_{i}^{\prime}(i=1,2)
\end{array}
&
\begin{array}
[c]{c}%
S^{2}\\
S^{2}%
\end{array}
&
\begin{array}
[c]{c}%
\text{{\small from }}B_{i}\text{, {\small not }}B_{3}\\
\text{{\small from }}B_{i}^{\prime}\text{, {\small not }}B_{3}^{\prime}%
\end{array}
&
\begin{array}
[c]{c}%
C_{i}^{-}=C_{i\text{ }}\\
C_{i}^{\prime-}=C_{i\text{ }}^{\prime}%
\end{array}
& \text{{\small spacelike}}%
\end{array}
\]

\begin{center}
Table 1. Main Features of the Submanifolds $N,B_{i}^{\prime},B_{i}%
,C_{i}^{\prime}$ and $C_{i}$
\end{center}

%

\begin{figure}
[ptb]
\begin{center}
\includegraphics[
height=5.3497in,
width=4.5766in
]%
{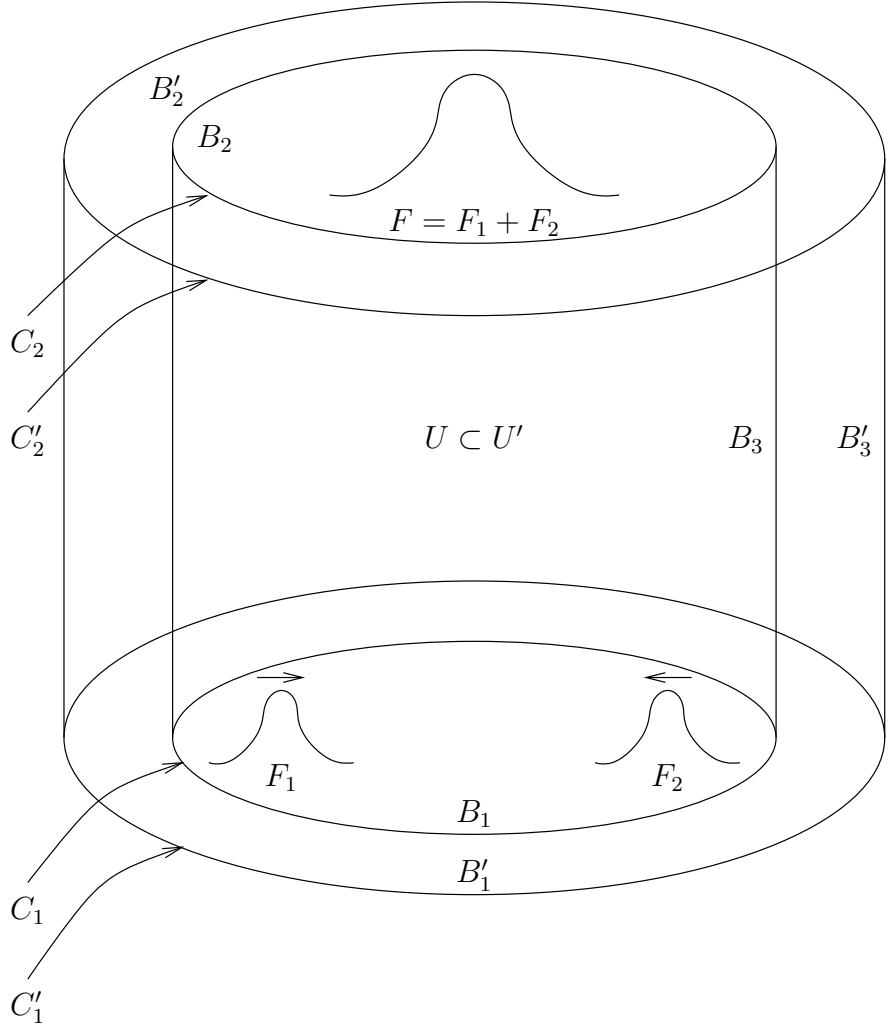}%
\caption{Standard Cylinder $N$ in Minkowski Spacetime, its Boundary Manifolds
and the Field Configurations $F_{1}(0,\mathbf{x})$, $F_{2}(0,\mathbf{x})$ and
$F(\mathfrak{t},\mathbf{x})=F_{1}(\mathfrak{t},\mathbf{x})+F_{2}%
(\mathfrak{t},\mathbf{x})$}%
\label{Figure 1}%
\end{center}
\end{figure}

\end{document}